\documentclass{PoS}
\usepackage{graphics}

\title{Masses of Black Holes in Active Galactic Nuclei:
Implications for NLS1s (invited)}

\ShortTitle{Black Hole Masses}

\author{\speaker{Bradley M. Peterson}\\
        Deparment of Astronomy and Center for Cosmology and
AstroParticle Physics, The Ohio State University\\
        E-mail: \email{peterson@astronomy.ohio-state.edu}}

\abstract{I review how AGN black hole masses are calculated
from emission-line reverberation-mapping data, with particular
attention to both assumptions and caveats. I discuss
the empirical relationship between AGN luminosity
and broad-line region radius that underpins the indirect
methods by which most AGN masses are estimated. I also
discuss how line widths are characterized in this method
and illustrate how different ways of measuring the line-widths
can lead to systematic errors in the mass scale.
I discuss specific implications for NLS1 galaxies and consider
whether the NLS1 phenomenon is better explained by 
source inclination or by Eddington rate, and conclude that
there is evidence that {\em both} of these effects are
contributing factors and that at least the high-Eddington
rate NLS1s are physically similar to some high-luminosity
quasars.}

\FullConference{Narrow-Line Seyfert 1 Galaxies and their place in the Universe - NLS1,\\
		April 04-06, 2011\\
		Milan Italy}

\begin{document}

\section{Introduction}

The masses of black holes in the nuclei of galaxies are measured
by observing how they accelerate test masses in their vicinity,
as is always the case in astronomy.
The ``test masses'' in the case of galactic nuclei are
stars or gas in the nuclear regions. 
The particular advantage of using stars as the test masses
around black holes is that they are subject only to gravitational
forces. Unfortunately, stellar dynamics can be modeled
accurately only with high angular-resolution spectroscopy;
the black hole radius of influence,
\begin{equation}
\label{eq:RBH}
R_{\rm BH} = 2 G M_{\rm BH}/\sigma_*^2,
\end{equation}
where $\sigma_*$ is the stellar velocity dispersion of 
the host galaxy bulge, must be resolved, or
at least nearly so.
Measurements of gas motions, on the other
hand, allow us to probe much closer to the black hole, on
scales far below the black hole radius of influence in the
case of reverberation mapping \cite{BMcK82,Peterson93}.
However, gas also responds to 
non-gravitational forces, such
as hydromagnetic acceleration and radiation pressure, which
must be accounted for.

Methods that rely on measurements of the motions of stars or gas that are
accelerated by the black hole are ``direct methods.'' These include
modeling of both stellar and gas dynamics and reverberation mapping.
We also employ ``indirect methods'' of mass determination
that measure observables that are {\em correlated} with black hole mass: these 
include the relationships between
black hole mass and stellar bulge velocity dispersion
(the $M_{\rm BH}$--$\sigma_*$ relationship
\cite{Ferrarese00, Gebhardt00a,Gultekin09}),
black hole mass and host-galaxy bulge luminosity
(the $M_{\rm BH}$--$L_{\rm bulge}$ relationship, which
is also sometimes known as the Magorrian \cite{Magorrian98}
relationship),
the fundamental plane, and AGN-specific relationships,
such as that between the AGN luminosity and the
size of the broad-line region (BLR), as discussed
below and by Bentz \cite{BentzPoS}. It is also useful to
distinguish among ``primary,'' ``secondary,'' and 
``tertiary'' methods of mass determination: primary
methods are those that require the 
fewest assumptions. Certainly, in the case of supermassive black holes,
mass measurements based on the proper motions and radial
velocities (Sgr A*) or megamasers (NGC 4258) are primary.
Also generally regarded as primary in the measurement
of nuclear supermassive black holes are stellar and gas
dynamics. Reverberation mapping, however, as it is 
currently practiced, requires an external calibration
of the zero-point for its mass scale through correlations
between the black hole mass and host-galaxy properties
that are assumed to be the same for the host galaxies
of active and quiescent black holes; for this reason
reverberation mapping is a ``secondary method,'' although it
is still a ``direct method.'' However, as discussed
below, reverberation mapping has the potential to become
a primary method. Since it will be the only method that
is potentially extendable to high redshift 
and the method most applicable to NLS1s, it
will be the main focus of the rest of this contribution.

\section{Results from Reverberation Mapping}

The desired outcome of reverberation mapping
is usually expressed in terms of a ``velocity--delay
map'' (also called the ``transfer function'' \cite{BMcK82})
$\Psi(\tau, V_{\rm LOS})$
which is the six-dimensional phase space of the
BLR (velocity field and geometry) projected into
the two observables, time delay $\tau$ and line-of-sight
(i.e., Doppler) velocity $V_{\rm LOS}$. The data
requirements for successful recovery of 
$\Psi(\tau, V_{\rm LOS})$ from spectrophotometric
monitoring are fairly stringent \cite{Horne04} and it
is therefore only recently that high-quality maps
are starting to appear \cite{Bentz10b,DenneyPoS}.
It is much more common to extract only the
mean response time for an emission line
integrated over all Doppler velocities. Simple
cross-correlation of the continuum and emission-line 
light curves gives a mean response time for
a particular emission line, and this mean response
time is weighted toward physical conditions where
both the emissivity and responsivity
(marginal change in emissivity in response to
a continuum change) are high.
Even in this simplified analysis, 
advances continue to be made as improved mathematical
descriptions of AGN variability \cite{Kelly09,Koz10}
have led to more observationally motivated
statistical modeling of the continuum behavior in the gaps in coverage
in the real time-series sampling. Recent improvements in
this methodology \cite{Zu11} allow self-consistent determination of the
uncertainties and, since multiple lines with different lags
can be dealt with simultaneously, make it possible to back-fill
some of the gaps in time series. The new methodology yields
lags that are largely consistent with those based on current
cross-correlation practices, but with reduced uncertainties
in the lag measurements and, indeed, more robust estimates
of these uncertainties.

At the time of this workshop, reverberation lag measurements have
been made for around 50 AGNs. 
Since the compilation of Peterson et al.\ \cite{Peterson04},
there have been new reverberation results from MDM Observatory
and Crimean Astrophysical Observatory
\cite{Bentz06b,Bentz07,Denney06,Denney09b,Denney09c,Denney10,Grier08,GrierPoS},
the Lick AGN Monitoring Program (LAMP) 
\cite{Bentz08, Bentz09c, Bentz10a, Bentz10b,Barth11}
and other programs \cite{Sergeev11,Stalin11}
that have concentrated primarily on the Balmer lines
in low-redshift AGNs. There have also been new measurements of the
C\,{\sc iv} $\lambda1549$ response in the
dwarf Seyfert NGC 4395 \cite{Peterson05}
and in one high-redshift quasar \cite{Kaspi07}.
The Mg\,{\sc ii} $\lambda2798$ response has been
measured reliably in only one case,
NGC 4151 \cite{Metzroth06}.

The major conclusions that can be drawn from these studies
are:
\begin{enumerate}
\item Different emission lines respond on different time scales,
with lines characteristic of the highest ionization gas responding
first, providing evidence for ionization stratification of the
BLR. The size of the BLR, as measured by a particular
emission-line time delay $\tau$, is inferred to be $R_{\rm BLR}
 = c \tau$.
\item In every case where the time delays for
multiple broad emission lines have been
measured \cite{PW99,PW00, Kollatschny03a,Bentz10a}
there is a tight anticorrelation between line width $\Delta V$
and time lag $\tau$ such that the product $\Delta V^2\,\tau$ is constant
(Figure 1), as expected if the dynamics of the BLR is
dominated by the gravitational potential of the central black hole.
Strictly speaking, it means that the BLR gas is responding to
an inverse-square force or forces, so radiation pressure might also play
a role \cite{Marconi08} (see Marconi's contribution to these proceedings
for a more thorough discussion \cite{MarconiPoS}).
\item The time lag for the emission-line response to
continuum variations is longer in more
luminous objects \cite{Kaspi00,Kaspi05}. After separating the 
AGN luminosity $L_{\rm AGN}$ from the starlight contamination by
the host galaxy \cite{Bentz06a,Bentz09b}, 
the relationship between these quantities is consistent
with $R_{\rm BLR} \propto L_{\rm AGN}^{1/2}$. This is well-established
only for H$\beta$, but the more limited reverberation results
for C\,{\sc iv} \cite{Peterson04,Peterson05,Kaspi07} are
consistent with the same relationship. This 
radius--luminosity ($R$--$L$) relationship
provides the underpinning for indirect methods of measuring
AGN black-hole masses over cosmic time \cite{BentzPoS,VestergaardPoS}.
\item In every case in which the response of a particular
emission line has been measured on multiple occasions, the
timescale for response of the emission line is
consistent with $\tau \propto L_{\rm UV}^{1/2}$,
where $L_{\rm UV}$ is the luminosity in the observable UV,
as close to the ionizing continuum as possible
 \cite{Peterson02}.
The emission-line width also changes in a way that
is at least approximately consistent with a constant
value of the product $\Delta V^2\,\tau$.
\end{enumerate}
\begin{figure}
\begin{center}
\includegraphics[scale=0.5]{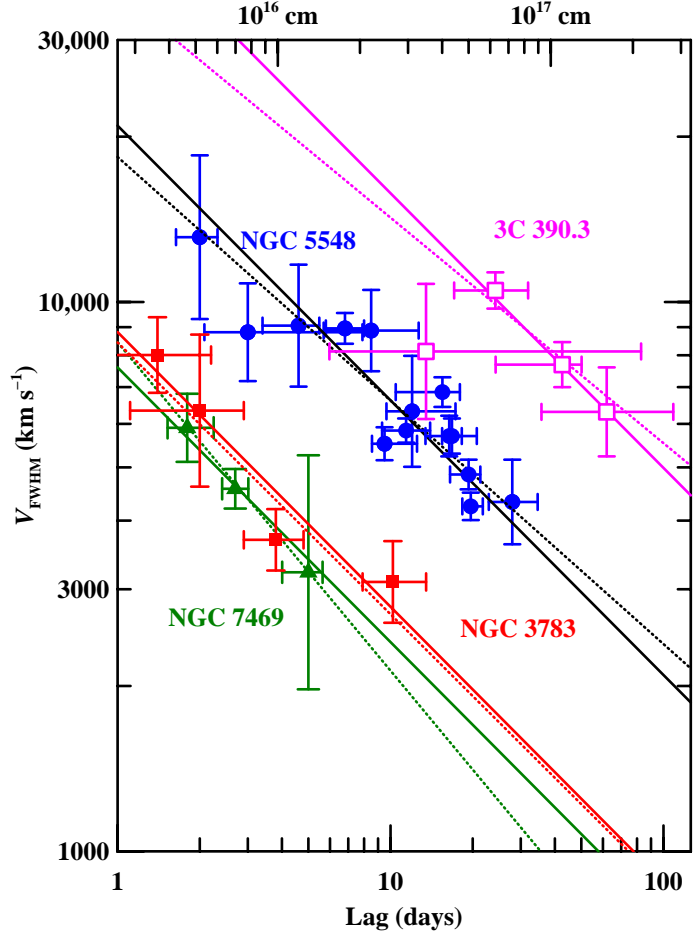}
\caption{Line-width vs.\ emission-line lag (in days)
for four reverberation-mapped AGNs. In each case,
the data are consistent with a virial relationship,
i.e., $V \propto \tau^{-1/2}$. The dotted lines are
show the best-fit slopes and the solid lines show
the best-fit virial relationship
\cite{PW99,PW00}.}
\end{center}
\end{figure}
It is worth mentioning that the
strong optical Fe\,{\sc ii} blends that are prominent features
of Seyfert 1 spectra are seen to vary over time scales that are
long compared to reverberation time scales, but do not appear
to vary much if at all on reverberation time scales
\cite{Vestergaard05,Kuehn08}. This may be because the emissivity and
responsivity distributions for these lines are not highly
localized, as they seem to be in the case of most of the
prominent emission lines in AGNs. In other words, these features
{\em vary}, but they do not {\em reverberate}.

Results (2) and (4) above suggest that we can measure the central
black hole mass by combining the emission-line width $\Delta V$
and time lag $\tau$, i.e.,
\begin{equation}
\label{eq:BHmass}
M_{\rm BH} = f \left( \frac{\Delta V^2 c\tau}{G} \right).
\end{equation}
The quantity in parentheses is based on the observable
quantities and has units of mass; we refer to it as
the ``virial product.'' The factor $f$ is a dimensionless
number of order unity into which we subsume all of our
ignorance of about the geometry and kinematics of the
BLR; what we are attempting to do is to characterize a
probably complex BLR with two observables that we hope will
somehow average over the complexity well enough to yield
a reasonable mass estimate. Let's consider the elements of
this equation individually:
\begin{itemize}
\item As noted earlier, $\tau$ is the mean time lag
for a given emission line. Practically speaking, it
is usually the centroid of the continuum/emission-line
cross-correlation function. It is unbiased as a function
of BLR inclination as long as the BLR is axisymmetric
and emits isotropically (and even when it does not 
emit isotropically, the correction factor is relatively
small compared to other projection effects). 
\item The line width $\Delta V$ is more problematic,
as there are multiple issues to consider. The first of
these is what line-width measure to use? Commonly used
measures are (a) full-width at half maximum (FWHM) and
(b) line dispersion $\sigma_{\rm line}$, 
the second moment of the line profile. Both measures have
liabilities: FWHM is more sensitive to both
random noise, unremoved narrow-line components,
and absorption due to intervening gas (at least in
the case of resonance lines like C\,{\sc iv}\,$\lambda1549$)
and $\sigma_{\rm line}$ is more sensitive to blending
with other features\footnote{We are currently experimenting
with using interpercentile widths \cite{Whittle85} as
an alternative characterization of line width,
as have Fine et al.\ \cite{Fine08},
but it is premature to conclude
anything at this point.}.
Second, given an array of spectra from a reverberation-mapping
monitoring campaign, we 
have two obvious alternatives for forming
the spectrum from which we measure the line widths.
The first of these is the mean spectrum, which we
construct by determining the average flux in
each wavelength bin of the spectrum over the duration
of the monitoring campaign. The second of these  is
an ``rms spectrum'' that is comprised
of the root-mean-square fluxes in each 
wavelength bin of the spectrum over the duration of
the campaign. The rms spectrum affords two particular
(related) advantages: (a) the rms spectrum isolates
the emission-line gas that is actually responding to
the continuum variations and omits the unchanging
parts (thus automatically excluding narrow-line components
and host-galaxy starlight), and (b) we find in practice that
the variable parts of the emission lines are much less
blended with other features, mostly because some features
such as the optical Fe\,{\sc ii} lines do not vary on
reverberation time scales, but partly because the widths
of the emission lines are typically narrower in the rms
spectrum than in the mean spectrum (possibly because the
very highest velocity BLR gas is optically thin
\cite{Ferland90,Shields95}).
A nice comparison of a mean and an rms spectrum is
shown in Grier et al.'s contribution to these proceedings \cite{GrierPoS}.
\item The scaling factor $f$ converts the convenient virial product,
based on the observables, to an actual mass. If the 
BLR is a flat Keplerian disk observed at inclination
$i$, the scaling factor would include a 
$1/\sin^2 i$ term to account for the velocity projection. It could
also include a correction factor for $\tau$ if the line
emission is anisotropic. Also --- and this is a critical
point that is often misunderstood --- the value of
$f$ depends on what line-width measure is being used.
This is an extremely important point that I will return to
below.
\end{itemize}
It should be clear from this discussion that the scale
factor $f$ is different for every AGN; indeed, in principle
it could be somewhat different for different emission lines
in the same AGN, but not too much different since, as
shown in Figure 1, the virial product is approximately constant
for all emission lines. Without additional information
on the kinematics, geometry, and inclination of the AGN,
we cannot determine $f$ for a particular source. 

We could also determine $f$ if we had an independent
measure of the black hole mass. Unfortunately, there are
very few cases where alternative direct measurements exist (see below),
and none of these are high precision measurements.
We can, however, use other {\em indirect} methods to estimate
the black hole masses in reverberation-mapped AGNs and
use these to determine the scale factor. However, since
the fidelity of individual mass measurements by indirect
methods is low, we can realistically hope to obtain only
a mean value $\langle f \rangle$ for the reverberation-mapped
sample and thus effect a statistical scale factor for the
reverberation-mapped sample. This has been done by assuming
that the  $M_{\rm BH}$--$\sigma_*$ relationship for AGNs
is the same as it is in quiescent galaxies. It is well-known
that reverberation-based black hole masses scale with
stellar bulge velocity dispersion in a fashion similar to that
seen in quiescent galaxies \cite{Gebhardt00b,Ferrarese01,
Onken04,Nelson04,Woo10} so it is not too much of a stretch
to assume that the zero-points for the
AGN and quiescent-galaxy 
are identical. The most recent analysis
gives a value of $\langle f \rangle = 5.25 \pm 1.21$ \cite{Woo10}
(though see \cite{Graham11} and \S{\ref{section:HGP}}).
The scatter around the AGN
$M_{\rm BH}$--$\sigma_*$ relationship is estimated to
be $\sim 0.4\, {\rm dex}$, which is thus an estimate of the
typical uncertainties in reverberation-based masses, assuming
that the intrinsic scatter in the $M_{\rm BH}$--$\sigma_*$ relationship
is much smaller than this (but more on this later).

The reverberation-based black hole masses in the literature
are computed from equation (\ref{eq:BHmass}) with
$\Delta V$ taken to be the line dispersion in the rms spectrum,
$\tau$ is the cross-correlation centroid, and
$f$ is the value based on the 
$M_{\rm BH}$--$\sigma_*$ relationship \cite{Onken04, Woo10}.
In a small number of cases, we can compare the 
reverberation-based masses with those from other direct methods.
Two cases are shown in Table 1, NGC 3227 and NGC 4151. In both
cases, the direct methods are in reasonable agreement, particularly
considering the factor of three or so systematic uncertainty
in the reverberation-based masses and probably comparable
systematic uncertainties in the other primary methods.
Another check on the accuracy of reverberation-based masses is to compare
the $M_{\rm BH}$--$L_{\rm bulge}$ relationship for AGNs to
that for quiescent galaxies \cite{Bentz09a}.
The agreement is found to be very good.

\begin{table}
\begin{center}
\begin{tabular}{lcc}
\multicolumn{3}{c}{{\bf Table 1:}
Black Hole Masses (Units of $10^6\,M_{\odot}$)}\\
\hline
Galaxy & NGC 3227 & NGC 4151 \\
\hline
{\em Direct methods:} \\
Stellar dynamics & $7$--$20$ \cite{Davies06} &
$< 70$ \cite{Onken07}\\
Gas dynamics & $20^{+10}_{-4}$ \cite{Hicks08} & 
$30^{+7.5}_{-22}$ \cite{Hicks08}\\
Reverberation & $7.63 \pm 1.7$ \cite{Denney10} & 
$46\pm5$ \cite{Bentz06b}\\
\hline
{\em Indirect methods:} \\
$M_{\rm BH}$--$\sigma_*$ & 25 & 6.1 \\
$R$--$L$ scaling & 15 & 65 \\
\hline
\end{tabular}
\end{center}
\end{table}

Use of a mean value $\langle f \rangle$
averages over the specific values of $f$ for
individual sources: if it is unbiased, 
$\langle f \rangle$ will result in equal numbers
of mass overestimates and underestimates, but
{\em these will not be a function of either time delay or line width.}

\section{Indirect Mass Measurements Anchored by Reverberation Mapping}
As noted above (result 3 in the last section), 
reverberation mapping has revealed a tight
relationship between the AGN continuum
luminosity and the radius of the BLR \cite{BentzPoS}
of the form
\begin{equation}
R_{\rm BLR} \propto L_{\rm AGN}^\alpha,
\end{equation}
where empirically we find $\alpha \approx 0.5$,
which is, as is often noted, consistent with
the theoretical expectation.
In the early days of photoionization equilibrium
modeling of the BLR, Davidson \cite{Davidson72}
and others began to recognize that photoionization
models could be parameterized by (a) the shape of
the ionizing continuum, (b) the elemental abundances,
(c) the particle density $n_{\rm H}$ and column density of 
the photoionized gas, and 
(d) an ionization parameter defined by
\begin{equation}
U = \frac{Q({\rm H})}{4 \pi R^2 c n_{\rm H}},
\end{equation}
where $R$ is the separation between the source
of ionizing photons and the photoionized gas
(i.e., the BLR radius)
and $Q({\rm H})$ is the number of ionizing
photons produced per second by the
photoionizing source, i.e.,
\begin{equation}
Q({\rm H}) = \int \frac{L_\nu}{h\nu} d\nu
\end{equation}
and the integral is over all hydrogen-ionizing
frequencies. To some low-order of approximation,
AGN spectra are self-similar over many orders
of magnitude in luminosity, which suggests that 
the BLRs of all AGNs are
characterized by similar values of $U$
and $n_{\rm H}$. This leads to the prediction that 
\begin{equation}
R_{\rm BLR} = 
\left( \frac{Q({\rm H})}{4 \pi c n_{\rm H} U} \right)^{1/2}
\propto L_{\rm AGN}^{1/2},
\end{equation}
if we can use $L_{\rm AGN}$, the luminosity at some
observable wavelength, as
a proxy for the ionizing luminosity.
We characterize this expectation as 
``na\"{\i}ve'' since some of the implicit
assumptions are quite approximate: we know 
that the shape of the ionizing continuum should
change with black hole mass and that the BLR is
complex and cannot be characterized by fixed
values of $U$ and $n_{\rm H}$ \cite{Baldwin95,Korista97}.
In the early photoionization equilibrium calculations in the 1970s,
the size of the BLR was not considered to be an
important parameter as the models suggested
that the BLR sizes in quasars would range from
one to a thousand light years or so
\cite{Krolik78, Davidson79}, which would be 
unresolvable in distant quasars in any event. I suspect
that Davidson does not get enough credit for 
his insight on the ionization parameter and the $R$--$L$
scaling relationship that follows from it
because he wrote the ionization
parameter in terms of the ionizing flux
($U \propto F_{\rm ion}/n_{\rm H}$) rather than the 
ionizing luminosity ($U \propto L_{\rm ion}/R^2 n_{\rm H}$),
as it was written later \cite{McKee75}, so the 
BLR radius did not appear explicitly.
Nevertheless, the prediction 
of the $R$--$L$ relationship was so well-known that 
a decade later it was often quoted without attribution 
(e.g., \cite{Mathews84}) when
the BLR size became an interesting parameter because the first
emission-line variability studies suggested that the BLR
was much smaller than predicted by the photoionization equilibrium
models \cite{Peterson85}. 
The $R$--$L$ relationship was looked for 
as a prediction of photoionization theory even
when the first crude reverberation lags
were measured \cite{Koratkar91}; it was
always understood, of course, that the central mass could be easily
estimated from a single quasar spectrum via
equation (\ref{eq:BHmass}) by using 
$R_{\rm BLR}$ from the $R$--$L$ relationship
and $\Delta V$ from an emission-line width measurement.
This ``photoionization method'' \cite{Wandel99, Kaspi00}
thus provides a significant shortcut to estimating
AGN black hole masses\footnote{This method for estimating
quasar masses has been sometimes, 
misleadingly in my opinion, referred to
as the ``Dibai method'' \cite{Bochkarev09}.
Dibai \cite{Dibai77} noted that since AGN emission lines
have the same equivalent widths regardless of luminosity
$L_{\rm line} \propto L_{\rm continum}$. If one then assumes
constant emission-line emissivity per unit volume $\epsilon$, then
$L_{\rm line} = \epsilon 4 \pi R_{\rm BLR}^3 /3$, or
$R_{\rm BLR} \propto L_{\rm cont}^{1/3}$ and 
$M_{\rm BH} \propto \Delta V^2 L_{\rm cont}^{1/3}$.
So while the predicted sizes for the BLRs in nearby
AGNs were reasonable, the physics is incorrect and the
functional form of the $R$--$L$ relationship is wrong.
Certainly Dibai did a
lot of important work that was admittedly often overlooked in
the West, but this incorrect
attribution overlooks the more correct physical insights provided by
Kris Davidson, Gordon MacAlpine, Chris McKee, Bruce Tarter,
and others.}.
The general methodology has been extended to other emission lines,
in particular, C\,{\sc iv}\,$\lambda1549$ 
\cite{Vestergaard02,Vestergaard04,Vestergaard06}
and Mg\,{\sc ii}\,$\lambda2798$ \cite{McLure02}.
Similar $R$--$L$ relationships using various
proxies for the continuum luminosity
have also been shown to be viable \cite{Greene10}.

The bottom rows of Table 1 show two indirect measurements
of the black hole masses in NGC 3227 and NGC 4151. The
indirect methods are in good agreement with the direct 
methods in the case of NGC 3227, but not quite as good
in the case of NGC 4151. But generally the picture holds
together to an accuracy of $\sim 0.5\,{\rm dex}$.

\section{Implications for NLS1s}

\subsection{Inclination or Eddington Ratio?}

There are two competing explanations for the properties
of NLS1s:
\begin{enumerate}
\item NLS1s might be low-inclination sources,
viewed nearly pole-on. If the
broad-line emission arises primarily in a rotating
Keplerian disk, for example, the projected rotational
speed is decreased by a factor of $\sin i$, thus accounting
for the unusually narrow broad lines.
\item NLS1s might be AGNs that are accreting at relatively
high Eddington ratio: for a given luminosity, objects
with the narrowest emission lines have the lowest black hole mass.
\end{enumerate}

There is certainly evidence that in general broad lines
widths are affected by inclination.
The widths of the Balmer lines
in core-dominated radio sources (those at low inclination)
are systematically lower than those in the spectra of lobe-dominated
(high-inclination) sources \cite{Wills86}. Similarly,
radio sources with flat spectral indices (low inclination
sources) have narrower line widths than radio sources
with steep spectral indices \cite{Jarvis06}. Of course, the fact
that ``broad'' lines are never narrower than 
$\sim 1000\,{\rm km\,s}^{-1}$ means that there must an an
axial component to the BLR velocity field, but it still seems
that the dominant motion might be rotational. 
Decarli et al.\ \cite{DecarliPoS} make a good case
that such a model can explain many of the properties
of NLS1s, although I don't believe inclination can explain
everything. Also, absorption-line and reverberation studies
\cite{CrenshawPoS, DenneyPoS}
both suggest that the well-known NLS1 NGC 4051 is indeed
observed at low-inclination.  And,
very importantly, some NLS1s have now been detected
in very high energies by {\em Fermi} (\cite{Abdo09,FoschiniPoS},
suggesting that we
are looking down the axis of these systems. 

Is inclination sufficient to account for the NLS1 phenomenon?
Probably not, in my opinion. I'll cite two specific examples.
One of the few radio-loud reverberation mapped AGNs is 
3C\,120, which has a relativistic jet with an inclination
$i < 20^{\rm o}$. The inclination correction 
to the mass is at least a factor of 10 for this inclination; but
3C\,120 nevertheless falls right on the 
$M_{\rm BH}$--$\sigma_*$ relationship rather than below it.
Mrk 110 is another interesting case, an NLS1 with an
independent black hole mass measurement based on the
gravitational redshift of the emission lines \cite{Kollatschny03b}.
As seen in Table 2, the reverberation mass is the largest
of the three available mass estimates: we would expect it
to be the {\em smallest} if the narrowness of the Balmer lines
is attributable to inclination only as the other two
measures should be independent of inclination. 

\begin{table}
\begin{center}
\begin{tabular}{lc}
\multicolumn{2}{c}{{\bf Table 2:} Black Hole Mass Measurements for Mrk 110  }\\
\hline
Method & Mass (Units of $10^6\,M_{\odot}$) \\
\hline
Reverberation: & $ 25 \pm 6$ \\
$M_{\rm BH}$--$\sigma_*$: & 4.8 \\
Gravitational redshift: & $14 \pm 3$ \\
\hline
\end{tabular}
\end{center}
\end{table}

Consider now the alternative explanation, that the
NLS1 phenomenon is a manifestation of low black-hole mass
at a given luminosity. Figure 2a shows the mass--luminosity
relationship for reverberation-mapped AGNs, with the
NLS1s highlighted. The NLS1s are generally high Eddington ratio
(i.e, accretion rate relative to the Eddington value)
objects, but they are also generally low-luminosity
sources. This, we shall see, is an accident of how the
NLS1 class was defined.

\begin{figure}
\begin{center}
\includegraphics[scale=0.5]{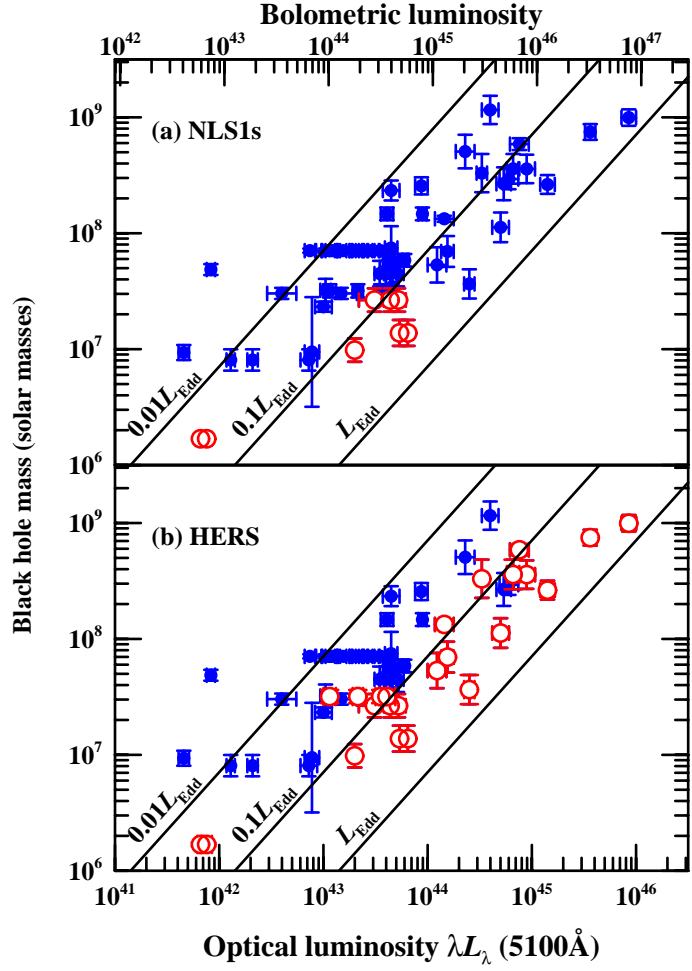}
\caption{$M_{\rm BH}$--$L_{\rm AGN}$ for the reverberation-mapped
AGNs. Lines of constant Eddington ratio are shown for
$\dot{m} = 0.01$, $\dot{m} = 0.10$ and $\dot{m} = 1.0$.
Multiple $L_{\rm AGN}$ values for a
fixed $M_{\rm BH}$ indicate multiple reverberation results
for some sources (notably NGC 5548 at $\sim 7 \times10^7\,M_\odot$).
The upper scale shows bolometric luminosity based on 
a nominal correction 
$L_{\rm bol} = 9\lambda L_{\rm AGN}$. In
the upper panel, NLS1s are shown as open cirles.
In the lower panel, high Eddington rate sources
(HERS) are shown as open circles.}
\end{center}
\end{figure}

Let's suppose that the common physics of the
NLS1 phenomenon is captured by the Eddington ratio.
The H$\beta$ line width is given by the virial equation
$\Delta V \propto \left( M_{\rm BH}/R_{\rm BLR} \right)^{1/2}$.
From the $R$--$L$ relationship of the last section
and the definition of the Eddington ratio
$\dot{m} = \dot{M}/\dot{M}_{\rm Edd} \propto \dot{M}/M_{\rm BH}$,
we see that the dependence of the line width is 
\begin{equation}
\Delta V \propto \left( \frac{M_{\rm BH}}{L^{1/2}_{\rm AGN}} \right)^{1/2} 
\propto \left( \frac{M_{\rm BH}}{\dot{M}^{1/2}} \right)^{1/2} 
\propto \left( \frac{M_{\rm BH}}{\dot{m}} \right)^{1/4}. 
\end{equation}
What this tells us is that for a fixed Eddington ratio $\dot{m}$,
the emission lines are broader for more massive black holes.
If this is correct, then we should be modifying our definition
somewhat, since higher mass black holes will have broader
lines even though the Eddington ratio is fixed 
\cite{Peterson07,DultzinPoS}.
Rather than modify the NLS1 definition, I'll just define (arbitrarily)
a larger class of ``high Eddington ratio sources (HERS)'' as those
with
\begin{equation}
\label{eq:HERSdefn}
\Delta V \leq \left( \frac{M_{\rm BH}}{10^7\,M_\odot} \right)^{1/4}
\ 2000\, {\rm km\ s}^{-1}.
\end{equation}
The HERS are highlighted in Figure 2b. If the NLS1s
are a high Eddington ratio phenomenon, then they are the
low-luminosity subset of the larger HERS class. But taken
as a group, they do not stand out from the other objects
in any properties except line width and profile and
the parameters derived directly from these parameters;
this is consistent with Bentz \cite{BentzPoS}, but
rather at odds with the results of other investigations
(see \S{\ref{section:HGP}}).

The bottom line is probably that NLS1s are yet another
``mixed bag'' (cf.\ \cite{Williams04})
of sources that include both high Eddington
rate accretors and low-inclination AGNs. Inclination effects
are clearly present, as are some NLS1 characteristics
that cannot be explained away as inclination effects.
Indeed, the case can be made that many NLS1s
are actually {\em both} pole-on {\em and} have high 
Eddington ratios \cite{BorosonPoS}.

\subsection{Characterizing the Line Width}

So far we have side-stepped the question of what is
the better measure of the line width, FWHM or
$\sigma_{\rm line}$? In this case, ``better'' means
which measure, if either, gives a black hole
mass that is unbiased. It is inescapable that at
least one of these measures
will introduce a bias into the mass scale because
the {\em shape} of AGN emission lines is
a function of their width. This is illustrated in
Figure 3, which shows ${\rm FWHM}/\sigma_{\rm line}$ as
a function of FWHM, where again we have highlighted
the HERS. Clearly the choice of which
measure we use for computing the masses is important.
This is illustrated very clearly in Figure 4,
which shows a direct comparison of black hole masses
based on FWHM with those based on $\sigma_{\rm line}$; 
the HERS have smaller masses for FWHM than
for $\sigma_{\rm line}$ and vice versa for the remainder
of the AGNs.

\begin{figure}
\begin{center}
\includegraphics[scale=0.5]{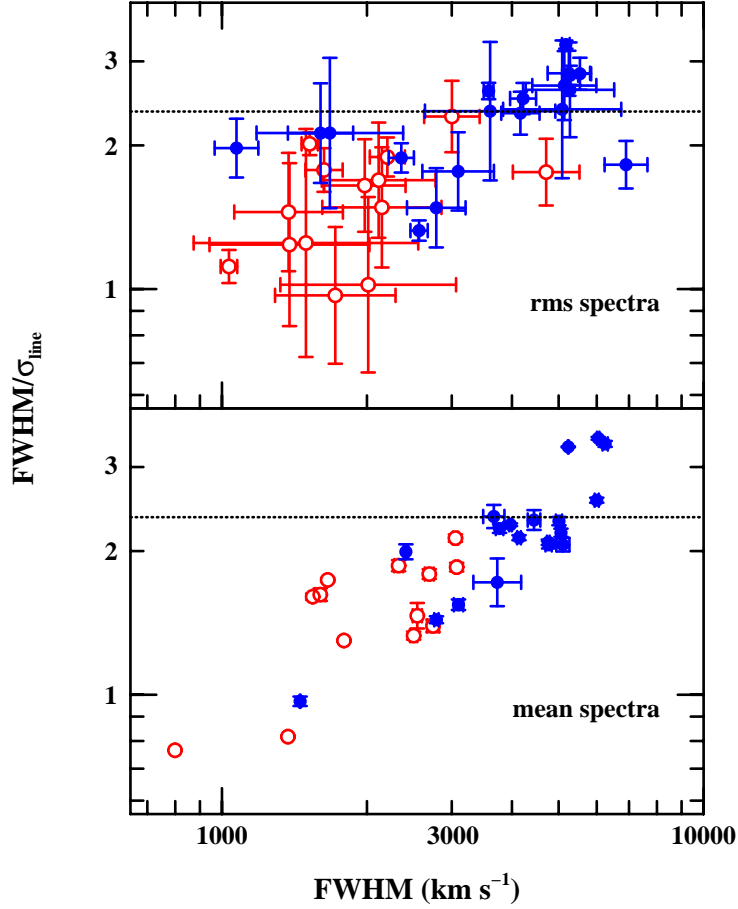}
\caption{The line-width ratio ${\rm FWHM}/\sigma_{\rm line}$ as
a function of FWHM for the H$\beta$ line in the
rms (top) and mean (bottom) spectra of reverberation-mapped AGNs.
In both panels, a dotted line shows the value for
a Gaussian profile, ${\rm FWHM}/\sigma_{\rm line} = 2.35$.
HERS are shown as open circles.}
\end{center}
\end{figure}

\begin{figure}
\begin{center}
\includegraphics[scale=0.5]{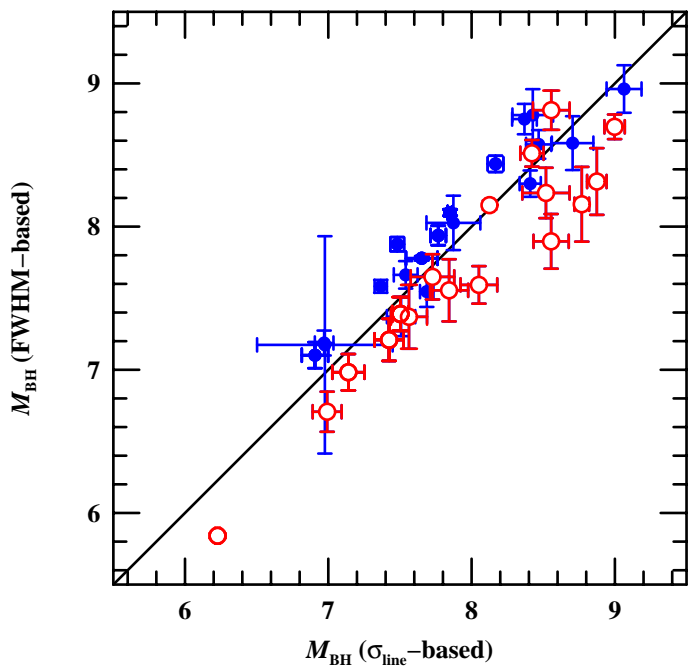}
\caption{A direct comparison of black hole masses
based on different line-width measures, FWHM
and the line dispersion $\sigma_{\rm line}$, in
both cases based on measurement of the H$\beta$ line in the
rms spectrum. HERS 
are shown as open circles. Relative to $\sigma_{\rm line}$,
FWHM yields lower masses for HERS because of the higher
kurtosis of the Balmer-line profiles in HERS spectra.}
\end{center}
\end{figure}

An argument can be made that $\sigma_{\rm line}$ is the
less-biased line-width measure, at least in the
case of lines measured in rms spectra
\cite{Peterson04, Collin06}, the variable part of
the spectrum. However, in individual spectra of AGNs,
generally the line wings are badly blended with other
features, and introduces systematic errors that are a function
of line width \cite{Denney09a}, but the errors introduced
are still smaller than the systematic differences that arise from
using FWHM instead of $\sigma_{\rm line}$.

The clear trend illustrated in Figure 3 demonstrates
that it is possible to correct FWHM to $\sigma_{\rm line}$;
this was, in fact, explicitly proposed by Collin et al.\
\cite{Collin06}. There are other formulations that have
suggested that the dependence on line width in equation
(\ref{eq:BHmass}) should be written as ${\rm FWHM}^\gamma$,
where $\gamma$ is a free parameter, which in practice
turns out to have a value slightly smaller than two
\cite{Wang09}; in essence, this is making the same correction
between FWHM and $\sigma_{\rm line}$, since the latter is
what was used to compute the reverberation-based masses
\cite{Peterson04}.

It is important to sort this out as incorrect parameterizations
will give misleading results. While the issue has not
been completely resolved at this point, there is 
a current controversy that in any case illustrates the
point. Steinhardt \& Elvis \cite{Steinhardt10} show
that if one plots black-hole mass versus AGN
luminosity for the SDSS DR5 quasars, in each redshift
bin it is found that while the lower-mass AGNs approach
the Eddington limit, the higher mass AGNs do not. In
other words, the axis of the distribution of quasars on the
$L_{\rm AGN}$--$M_{\rm BH}$ diagram (similar to Figure 2,
with the axes interchanged), does not run parallel to
lines of constant $\dot{m}$, but is rather more shallow.
The absence of higher mass AGNs with high values of the Eddington
ratio is referred to as the ``sub-Eddington boundary.''
The mass measurements used by Steinhardt \& Elvis are from
Shen et al.\ \cite{Shen08}, in which the
emission lines were fit with a pair of Gaussians,
one for the narrow component and one for the broad component,
and the black hole mass was computed using 
equation (\ref{eq:BHmass}) with the FWHM measurement
for the broad-component best-fit Gaussian.
In an independent study,
Rafiee \& Hall \cite{Rafiee11} went back to the 
original DR5 spectra and measured $\sigma_{\rm line}$ from
the original data. When they use $\sigma_{\rm line}$ 
to compute the black hole masses, they
find that the sub-Eddington
boundary disappears, and that in each redshift range
the distribution of AGNs in the 
$L_{\rm AGN}$--$M_{\rm BH}$ diagram  is along constant
values of $\dot{m}$. This, of course, does not
prove that the sub-Eddington boundary is an artifact
of how the line widths were characterized, but it
is certainly suggestive since the Rafiee \& Hall
result seems intuitively more likely to be correct.
Indeed, Figures 3 and 4 show
that using FWHM instead of $\sigma_{\rm line}$ will
certainly lead to overestimating the masses of
AGNs with the broadest lines and underestimating the
masses of the AGNs with the narrowest lines, which
would produce exactly the sub-Eddington boundary phenomenon.
A Gaussian fit ({\em rarely}
a good description of a broad-line profile) will
have the same effect (see Figure 3), so it's not
immediately clear whether the problem is principally
the Gaussian assumption or the use of FWHM.

Is the sub-Eddington boundary real? Perhaps. 
But I doubt it. 

\subsection{Host-Galaxy Properties}
\label{section:HGP}
Currently the reverberation-based  mass scale relies heavily
on correlations between black hole mass and host-galaxy
properties. It is explicitly assumed that these
relationships are the same in active and quiescent
galaxies. A particular concern is the nature of the
$M_{\rm BH}$--$\sigma_*$ relationship, which currently
provides the zero-point calibration for the
reverberation mass scale. Batcheldor \cite{Batcheldor10}
has argued that the $M_{\rm BH}$--$\sigma_*$ relationship
is a selection effect caused by preferentially observing sources
with a resolvable radius of influence (equation \ref{eq:RBH}),
though G{\"u}ltekin et al.\ \cite{Gultekin11} disagree.
Indeed, the existence of an AGN
$M_{\rm BH}$--$\sigma_*$ relationship also argues against this
as a selection effect,
since there is only a subtle radius-of-influence bias
in the AGN sample\footnote{Specifically, the most luminous 
objects in the reverberation database were selected at least in part
for their apparent brightness, so there is a bias
toward HERS, as is well-known for the Palomar--Green
sample from which these are drawn.}.

There is also active discussion about the host galaxies
of NLS1s and whether or not there are differences between
these and the hosts of other AGNs. Crenshaw, Kraemer, \& Gabel
\cite{Crenshaw03} claimed that NLS1 hosts are disproportionately
barred galaxies, and larger samples seem to confirm 
this \cite{Otha07}. This raises concerns as there is
evidence \cite{Graham11} that the $M_{\rm BH}$--$\sigma_*$ relationship
is different for barred galaxies than for non-barred and
elliptical galaxies. 
There are also indications that
the hosts of NLS1s have pseudo-bulges rather than classical
bulges \cite{Mathur11, Orban11, MathurPoS, OrbanPoS}, and
pseudo-bulges might \cite{BentzPoS} or might not \cite{Hu07}
not follow the $M_{\rm BH}$--$\sigma_*$ relationship.
This is a potential problem not only
for the overall reverberation-mapping mass scale, but 
for the {\em relative} masses of NLS1s compared 
to other AGNs.

\subsection{Toward Higher Precision Masses}

At the present time, the masses that can be obtained
from single-epoch spectra are probably accurate to 
about the $0.5\,{\rm dex}$ level, but we start to see
hints of bias in the mass scale at this point.
Proper characterization of line widths in 
individual (single-epoch)
spectra is probably the single most significant problem,
although this does not seem to be generally appreciated.
Beyond this, there are several important questions that
need to be addressed as we strive to make better black
hole mass measurements, and some of these are addressed in
other contributions to these proceedings:
\begin{itemize}
\item Are black hole mass/host-galaxy relationships the
same in AGNs and quiescent galaxies? Do AGNs and
quiescent galaxies differ in incidence of pseudo-bulges,
and how does this relate to black hole growth
and star-formation rates \cite{Sani10, SaniPoS}?
\item Are black hole mass/host-galaxy relationships the
same in NLS1s (or HERS) as in other AGNs? 
\item How reliable are the scaling relationships over
luminosity and redshift?
\item Does failure to account for radiation pressure
lead us to underestimate $M_{\rm BH}$? NLS1s (or HERS)
are the likely testing ground for this because the
radiation pressure term in the mass equation \cite{MarconiPoS} will
be relatively highest in these sources.
\item What problems do we encounter by using different
emission lines to determine $M_{\rm BH}$? In particular,
there are lingering doubts about the use of C\,{\sc iv},
despite good arguments that it is generally 
a good mass-indicator \cite{DenneyPoS,VestergaardPoS}.
My own observation (see also \cite{DenneyPoS,VestergaardPoS})
is that much of the evidence that
C\,{\sc iv} is problematic is based on data that are too poor
to address the issue.
\item How can we resolve the inclination/Eddington ratio
ambiguity in NLS1s (or HERS)?
\end{itemize}
Which brings us \ldots

\section{Back to Reverberation Mapping}

While we  expect that ongoing research 
will lead toward higher accuracy and higher precision black hole
masses, improvements in reverberation mapping is the direct 
path to address all of our concerns about black hole masses.
The velocity--delay map for the Balmer lines in Arp 151
\cite{Bentz10b} indicates a rich, complex structure and
a preliminary velocity--delay map for the NLS1 NGC 4051 \cite{DenneyPoS}
suggests that the Balmer-line emitting region in this source is a nearly
face-on disk. There are now several data sets of comparable
quality \cite{GrierPoS}, so additional
velocity--delay maps can be expected in the near future.
As more and better velocity--delay maps become available, it will be possible
to model the kinematics and geometry of the BLR 
directly from the reverberation data and thus
determine the central black hole masses without recourse
to other methods; this is what will be required for
reverberation mapping to become a primary method.

\begin{figure}
\begin{center}
\includegraphics[scale=0.5]{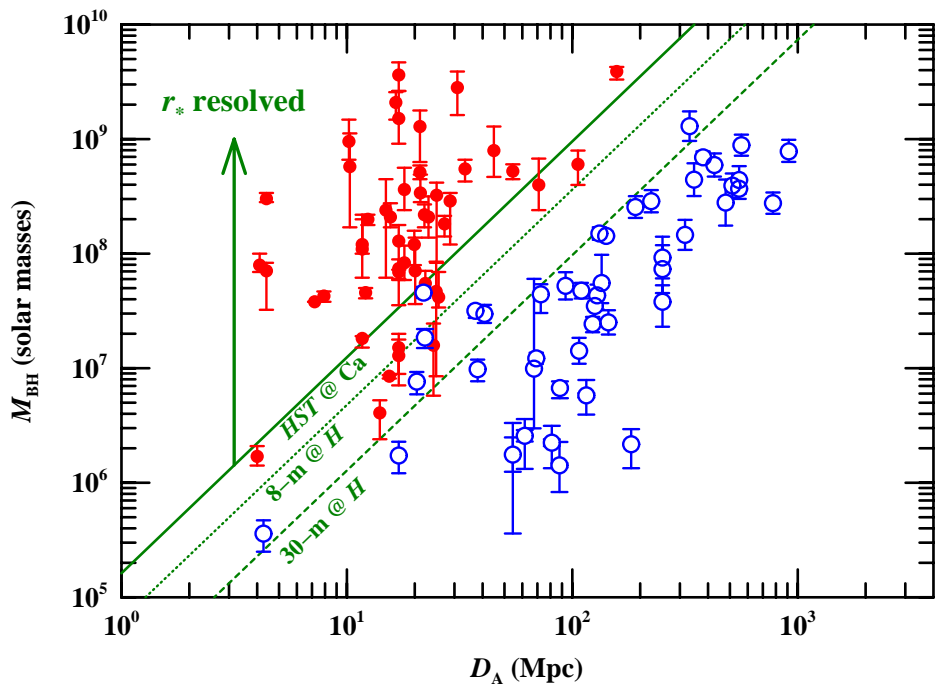}
\caption{$M_{\rm BH}$ versus luminosity distance for
nuclear black holes with direct measurements of their
masses. Reverberation-based masses are shown as open
circles and masses measured by stellar or gas dynamics
are shown as filled circles. The limiting black hole
radius of influence as a function of mass is shown
for the cases of {\em Hubble Space Telescope} at the
Ca\,{\sc ii} triplet ($\sim 8600$\,\AA), an 8-m
telescope at $H$-band, and a 30-m telescope at $H$-band.
A bare handful of AGNs are close enough to resolve
the radius of influence and thus measure their black
hole masses by the same techniques used on relatively
local galaxies.}
\end{center}
\end{figure}

The problem with reverberation mapping is, of course, simply
that it is both expensive (in terms of telescope time) and
risky (since the pattern of variability in a given source
is unpredictable).  In the future, it will probably
be possible to employ more clever strategies for reverberation
campaigns, for example, using photometric monitoring from
synoptic surveys like LSST to trigger campaigns at opportune
times. And given good photometric coverage, spectroscopic
monitoring can be sparser and tailored to the goals
of the program (for  
a good example of an opportunistic program, see \cite{Barth11}).

But it remains true that reverberation mapping is currently
the {\em only} way to measure the masses of black holes
in galaxies out to cosmologically interesing distances
and the only way to probe the low-mass end of the distribution, 
as illustrated in Figure 5. Even with a diffraction-limited
30-m telescope at $H$-band, it is impossible to resolve
the black hole radius of influence for 
$M_{\rm BH} < 10^8\,M_\odot$ at the distance of the Coma cluster.
Reverberation mapping, by substituting time resolution for
angular resolution, can extend direct measurements 
of black hole masses to
much larger distances.

\section*{Acknowledgments}
I am grateful for support of this research by grant at
The Ohio State University by the
NSF through grant AST-1008882 and by
NASA through grants HST-GO-11661 and HST-AR-12149
from the Space Telescope Science Institute. I thank
the organizers for an enjoyable and scientifically
productive meeting. My thanks go to K.~D. Denney, 
C.~J. Grier, R.~W. Pogge, 
and M. Vestergaard for comments on the draft
manuscript.

\end{document}